%% file: main.tex
\pdfoutput=1
\documentclass[sigconf]{acmart}
\acmConference[ESEC/FSE 2017]{11th Joint Meeting of the European Software Engineering Conference and the ACM SIGSOFT Symposium on the Foundations of Software Engineering}{4--8 September, 2017}{Paderborn, Germany}
\usepackage[framemethod=tikz]{mdframed}
\usepackage{lipsum}
\usetikzlibrary{shadows}
\usepackage{mathtools}
\usepackage{graphics}
\newmdenv[
tikzsetting= {fill=gray!10},
linewidth=1pt,
roundcorner=2pt, 
shadow=false
]{myshadowbox}
\usepackage{colortbl} 
\usepackage{blindtext, graphicx}
\usepackage{textcomp}
\usepackage{mathtools}
\usepackage{amsmath}
\usepackage{amsfonts}
\usepackage{color}
\usepackage{listings}
\usepackage{graphicx} 
\usepackage{multirow}
\usepackage{balance}
\usepackage{picture}
\usepackage{relsize}
\usepackage{soul}
\DeclarePairedDelimiter\abs{\lvert}{\rvert}%
\usepackage{array}
\usepackage{makecell}

\usepackage{verbatim}
\usepackage{algorithm}
\usepackage{algorithmicx}
\usepackage{algpseudocode}
\usepackage[export]{adjustbox}
\usepackage{mathtools}
\renewcommand{\footnotesize}{\scriptsize}
\definecolor{lightgray}{gray}{0.8}
\definecolor{darkgray}{gray}{0.6}

\definecolor{Gray}{rgb}{0.88,1,1}
\definecolor{Gray}{gray}{0.85}
\definecolor{Blue}{RGB}{0,29,193}
\usepackage{booktabs}
\usepackage{xspace}
\usepackage{todonotes}

\definecolor{MyDarkBlue}{rgb}{0,0.08,0.45} 
\newcommand{\quart}[3]{\begin{picture}(100,6)
{\color{black}\put(#3,3){\circle*{4}}\put(#1,3){\line(1,0){#2}}}\end{picture}}
\newcommand{\quartex}[3]{
\begin{picture}(13,6)
    {
        \color{black}
        \put(#3,3)
        {\circle*{4}}
        \put(#1,3)
        {\line(1,0){#2}}
    }
\end{picture}
}

\usepackage{soul}
\usepackage{color}

\definecolor{awesome}{rgb}{1.0, 0.13, 0.32}

\usepackage{amsmath}
\definecolor{Gray}{gray}{0.95}
\definecolor{LightGray}{gray}{0.975}
\definecolor{Mygreen}{HTML}{228B22}
\definecolor{Myred}{HTML}{800000}
\DeclareRobustCommand{\hlgreen}[1]{{\sethlcolor{Mygreen}\textcolor{white}{\hl{#1}}}}
\DeclareRobustCommand{\hlyellow}[1]{{\sethlcolor{yellow}\hl{#1}}}
\DeclareRobustCommand{\hlred}[1]{{\sethlcolor{Myred}\textbf{\textcolor{white}{\hl{#1}}}}}
\newcommand{\bi}{\begin{itemize}}
\newcommand{\ei}{\end{itemize}}
\newcommand{\be}{\begin{enumerate}}
\newcommand{\ee}{\end{enumerate}}

\newcommand{\fig}[1]{Figure~\ref{fig:#1}}

\usepackage{siunitx}

\newcommand{\custriangle}{\includegraphics[scale=0.5]{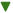}}
\newcommand{\cuscross}{\includegraphics[scale=0.4]{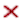}}
\newcommand{\cusdot}{\includegraphics[scale=0.5]{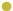}}

\usepackage{url}

\setcopyright{none}
\acmDOI{10.1145/3106237.3106238}
\acmISBN{978-1-4503-5105-8/17/09}
\acmConference[ESEC/FSE'17]{2017 11th Joint Meeting of the European Software Engineering Conference and the ACM SIGSOFT Symposium on the Foundations of Software Engineering}{September 4-8, 2017}{Paderborn, Germany} 
\acmYear{2017}
\copyrightyear{2017}
\acmPrice{15.00}

\begin{document}
\copyrightyear{2017} 
\acmYear{2017} 
\setcopyright{acmlicensed}
\acmConference{ESEC/FSE'17}{September 04-08, 2017}{Paderborn, Germany}\acmPrice{15.00}\acmDOI{10.1145/3106237.3106238}
\acmISBN{978-1-4503-5105-8/17/09}
%
\title{Using Bad Learners to find Good Configurations}


\author{Vivek Nair}
\affiliation{%
  \institution{North Carolina State University}  \city{Raleigh}
  \state{North Carolina, USA}}

\author{Tim Menzies}
\affiliation{%
  \institution{North Carolina State University}  \city{Raleigh}
  \state{North Carolina, USA}}

\author{Norbert Siegmund}
\affiliation{
  \institution{ Bauhaus-University }
  \city{Weimar}
  \state{Germany}}

\author{Sven Apel}
\affiliation{%
  \institution{University of Passau }
  \city{Passau}
  \state{Germany}}

\begin{abstract}

 Finding the optimally performing configuration of a software system for a given setting is often challenging. Recent approaches address this challenge by learning performance models based on a sample set of configurations.
 However, building an accurate performance model can be very expensive (and is often infeasible in practice). 
The central  insight of this paper is that   
exact performance values (e.g., the response time of a software system) are not required to rank  configurations and to identify the optimal one. 
As shown by our experiments, performance models that are cheap to learn but inaccurate (with respect to the difference between actual and predicted performance) can still be used rank configurations and hence find the optimal configuration. This novel \emph{rank-based approach} allows us to significantly reduce the cost (in terms of number of measurements of sample configuration) as well as the time required to build performance models. We evaluate our approach with 21 scenarios based on 9 software systems and demonstrate that our approach is beneficial in 16 scenarios; for the remaining 5 scenarios, an accurate model can be built by using very few samples anyway, without the need for a rank-based approach.
\end{abstract}

\keywords{
Performance Prediction, SBSE, Sampling, Rank-based method}

\maketitle

\section{Introduction}
This paper proposes an improvement of recent papers presented at ICSE'12, ASE'13, and ASE'15, which predict system performance based on learning influences of individual configuration options and combinations of thereof~\cite{siegmund2012predicting,guo2013variability,sarkar2015cost}. The idea is to measure a few configurations of a configurable software system and to make statements about the performance of its other configurations. Thereby, the goal is to predict the performance of a given configuration as accurate as possible.
We show that, if we (slightly) relax the
question we ask, we can build useful predictors using very small sample sets. Specifically,
instead of asking ``How long will this configuration run?'', we
ask instead ``Will \underline{this} configuration run faster than \underline{that} configuration?'' or ``Which is the \underline{fastest} configuration?''.

This is an important area of research since understanding system configurations
has become a major problem in modern software systems. 
In their recent paper, Xu et al.\ documented the  difficulties developers face
with understanding  the configuration spaces of their systems~\cite{xu2015hey}. As a result, developers tend to ignore over 83\% of configuration options, which leaves considerable optimization potential untapped and induces major economic cost~\cite{xu2015hey}.

With many configurations available for today's software systems, it is challenging to optimize for functional and non-functional properties. For functional properties, Chen et. al~\cite{chen2016sampling} and Sayyad et. al~\cite{sayyad2013scalable} developed fast techniques to find near-optimal configurations by solving a five-goal optimization problem. Henard et. al~\cite{henard2015combining} used a SAT solver along with Multi-Objective Evolutionary Algorithms to repair invalid mutants found during the search process. 

\begin{figure}[t]
\centering
\includegraphics[scale=0.35]{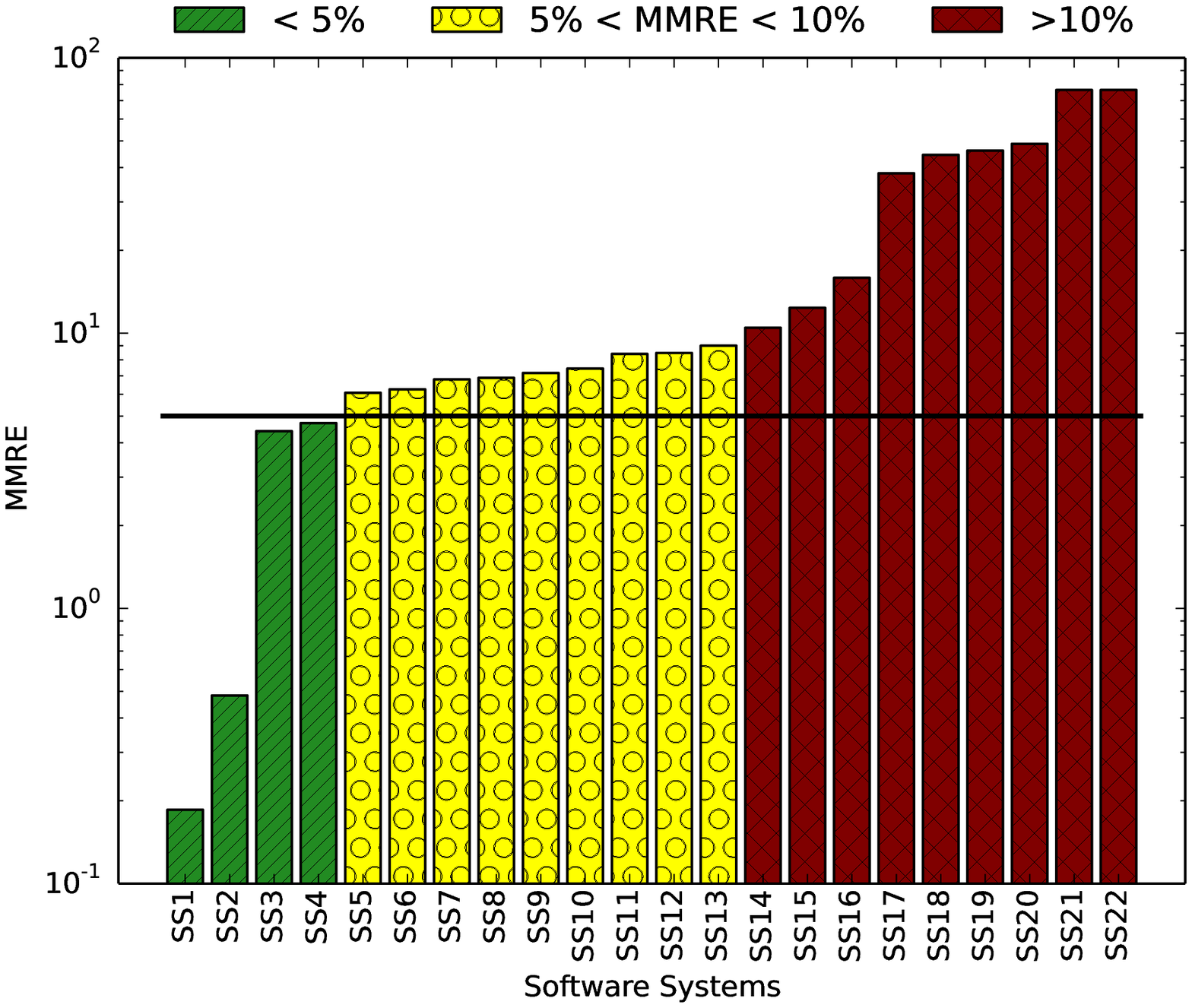}
\caption{{\small Errors of the predictions made by using CART, a machine learning technique (refer to Section~\ref{sec:systems}), to model different
software systems. Due to the results of \fig{learning_curve}, we use  30\%/70\% of the  valid configurations (chosen randomly) to train/test the model. }
}
\label{fig:model_efficiency}
\end{figure}

For non-functional properties, researchers have also developed a number of approaches. For example, it has been shown that the runtime of
a configuration can be predicted with high accuracy by sampling and learning performance models~\cite{siegmund2012predicting,guo2013variability,sarkar2015cost}. 
State-of-the-art techniques rely on configuration data from which it is possible to build very accurate models. For example, prior work~\cite{nair2017faster} has used sub-sampling to build predictors for configuration runtimes using predictors with error
rates less than 5\% (quantified in terms of \emph{residual-based} measures such as Mean Magnitude of Relative Error, or  MMRE, $(\sum_i^n  (|a_i - p_i|/a_i))/n$ where $a_i,p_i$ are the
{\em actual} and {\em predicted values}).
Figure~\ref{fig:model_efficiency} shows in \hlgreen{green} a number of real-world systems 
whose performance behavior can be modelled with high accuracy using state-of-the-art techniques.

Recently, we have come across software systems whose configuration spaces
are far more complicated and hard to model. For example, when the state-of-the-art technique of Guo at al.~\cite{guo2013variability} is applied to these software systems, the error
rates of the generated predictor is up to 80\%---see the 
\hlyellow{yellow} and
\hlred{red} systems of Figure~\ref{fig:model_efficiency}. The existence
of these harder-to-model systems raises a serious validity question 
for all prior research in this area:
\begin{itemize}
\item Was prior research merely solving easy problems?
\item Can we learn predictors for non-functional properties of more complex systems?
\end{itemize}
One pragmatic issue that complicates answering these two questions is the {\em minimal sampling problem}. It can be prohibitively expensive to run and test all configurations
of modern software systems since their configuration spaces are very large. For example, to obtain the data used in our experiments, we required over a month of  CPU time for measuring
(and much longer, if we also count the time required for compiling the code prior to
execution). Other researchers have commented that, in real-world scenarios, the cost
of acquiring the optimal configuration is overly expensive and time-consuming~\cite{weiss2008maximizing}. Hence, the goal of this paper must be:
\begin{enumerate}
\item Find  predictors for non-functional properties for the hard-to-model systems of Figure~\ref{fig:model_efficiency}, where learning accurate performance models is expensive.
\item Use as few sample configurations as possible.
\end{enumerate}

The key result of this paper is that, even when {\em residual-based} performance models are
inaccurate, {\em ranking} performance models can still be very useful for configuration optimization. Note that:
\begin{itemize}
\item Predictive models return a value for a configuration;
\item Ranking models rank $N$ configurations from ``best'' to ``worst''.
\end{itemize}
There are two very practical cases where such ranking models would suffice:
\begin{itemize}
\item Developers want to know the fastest configuration;
\item Developers are debating alternate configurations and want to know
which might run faster.
\end{itemize}
In this paper, we explore two research questions about  constructing ranking models.

  \noindent
  {\em  {\bf RQ1:}  Can  inaccurate predictive  models still accurately rank
configurations?}

We show below that,   even if a model has, overall, a low predictive accuracy (i.e., a high MMRE), the
predictions can still be used to effectively rank configurations.
The rankings are heuristic in nature and hence may be slightly inaccurate (w.r.t the actual performance value).
That said, overall, our rankings are surprisingly accurate. For example,
when exploring the configuration space of SQLite, our 
rankings are usually wrong only by less than 6 neighboring configurations -- which is a very small number
considering that SQLite has almost 4 million configurations.

\noindent {\em {\bf RQ2:} How expensive is a rank-based approach (in terms of
 how many configurations must be executed)?}

To answer this question, we studied the configurations of 21 scenarios based on 9 open-source systems.
We measure the benefit of our rank-based approach
as  the percentage of required measurements
needed by state-of-the-art techniques in this field (see Sarkar et al.~\cite{sarkar2015cost} presented at ASE'15). Those percentages were as follows -- 
Note that {\em lower} values are {\em better} and values under 100\% denote
an improvement over  the state-of-the-art (from Figure~\ref{fig:evals_ratio}):  

\begin{center}
\{5,5,5,5,5,10,20,20,20,20,30,30,35,35,40,40,50,50,70,80,80,110\}\%
\end{center}

\noindent
That is, the novel rank-based approach described in this paper is rarely worse than the  state of the art and often far better. For example, as shown later in Figure~\ref{fig:evals_ratio}, for one of the scenarios of Apache Storm, SS11, the rank-based approach uses only 5\% of the measurements used by a residual-based approach.

The rest of this paper is structured as follows: we first formally describe the prediction problem. Then, we describe the state-of-the-art approach proposed by Sarkar et al.~\cite{sarkar2015cost}, henceforth referred to as residual-based approach, followed by the description of our rank-based approach. Then, the subject systems used in the paper are described followed by our evaluation. The paper ends with a discussion on why a rank-based approach works; finally, we conclude.
To assist other researchers, a reproduction package with all our scripts and data are available
on GitHub.\footnote{
\url{https://github.com/ai-se/Reimplement/tree/cleaned_version}
}

\begin{figure}[t]
\centering
\includegraphics[scale=0.23]{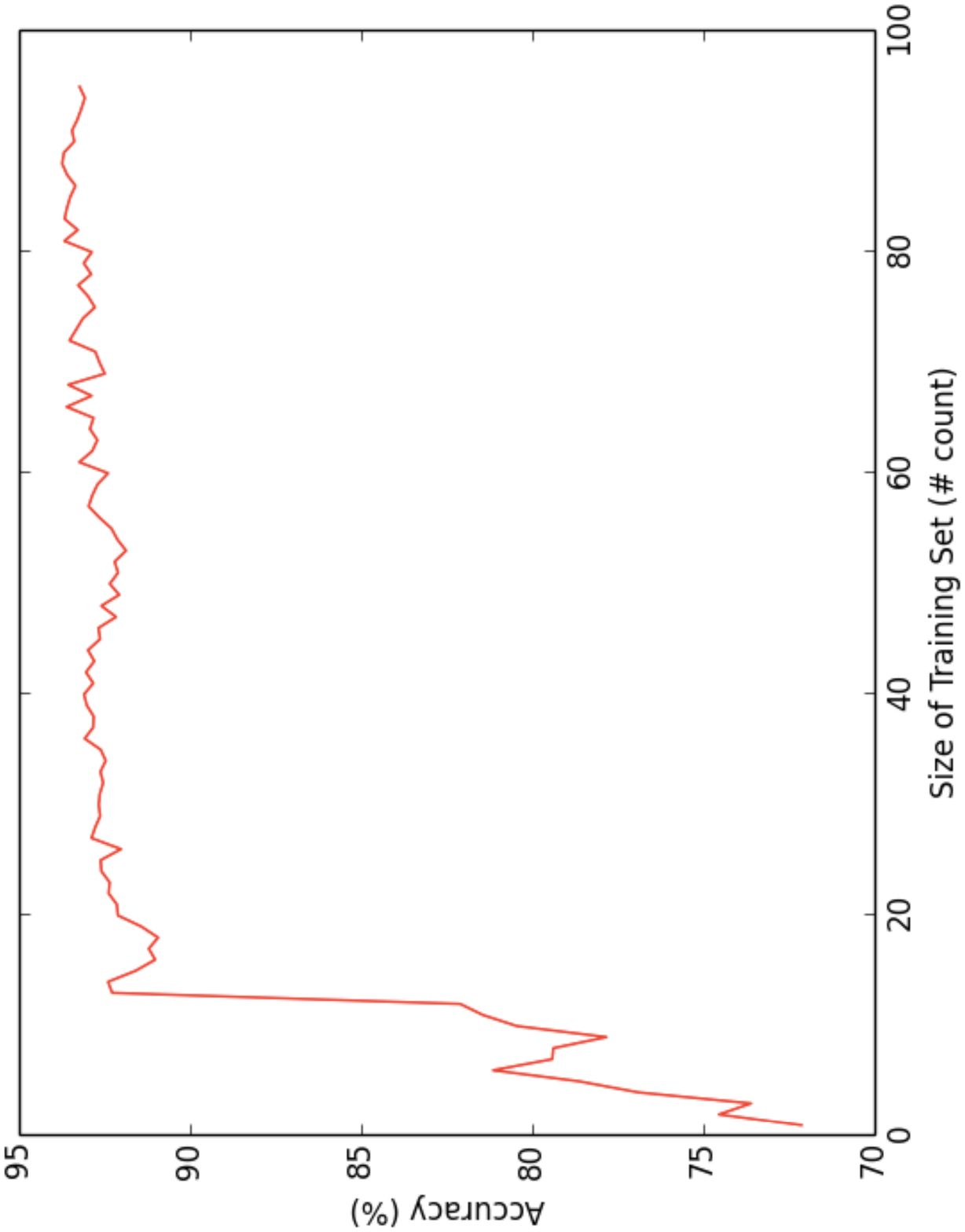}
\caption{{\small The relationship between the accuracy (in terms of MMRE) and the number of samples used to train the performance model of the Apache Web Server. Note that the accuracy does not improve substantially after 20 sample configurations.}
}
\label{fig:learning_curve}
\end{figure}

\begin{figure}[t]
\small
\hspace{0.2cm}\begin{lstlisting}[xleftmargin=5.0ex,mathescape,frame=none,numbers=left]
  # Progressive Sampling
  def progressive(training, testing, lives=3): 
    # For stopping criterion
    last_score = -1
    independent_vals = list()
    dependent_vals = list()
    for count in range(1, len(training)):    
      # Add one configuration to the training set
      independent_vals += training[count]      
      # Measure the performance value for the newly
      # added configuration 
      dependent_vals += measure(training_set[count])  
      # Build model
      model = build_model(independent_vals, dependent_vals)      
      # Test Model
      perf_score = test_model(model, testing, measure(testing))
      # If current accuracy score is not better than
      # the previous accuracy score, then loose life
      if perf_score <= last_score:
        lives -= 1
      last_score = perf_score
      # If all lives are lost, exit loop
      if lives == 0: break 
    return model
\end{lstlisting}
\caption{\small{Pseudocode of progressive sampling.}}
\label{fig:progressive_sampling}  
\end{figure}

\section{Problem Formalization}\label{sec:problem_formal}

A configurable software system has a set $X$ of configurations $x \in X$. Let $x_i$ indicate the \textit{ith} configuration option of configuration $x$, which takes the values from a finite domain $Dom(x_i)$. In general, $x_i$ indicates either an (i) integer variable or a (ii) Boolean variable. The configuration space is thus $Dom(x_1) \times Dom(x_2) \times ... \times Dom(x_n)$, which is the Cartesian product of the domains, where $n = |x|$ is the number of configuration options of the system. Each configuration ($x$) has a corresponding performance measure $y \in Y$ associated with it. The performance measure is also referred to as dependent variable. We denote the performance measure associated with a given configuration by $y=f(x)$.  We consider the problem of ranking  configurations ($x^*$) that such that $f(x)$ is less than other configurations in the configuration space of $X$ with few measurements.
\begin{equation}
    f(x^*) \le f(x),~~~ \forall x \in {X\setminus x^*}
\end{equation}
 
 Our goal is to find the (near) optimal configuration of a system where it is not possible to build an accurate performance model as prescribed in earlier work.

\section{Residual-based Approaches}\label{sec:residual}

In this section, we discuss the residual-based approaches for building performance models for configurable software systems. For further details, we refer to Sarkar et. al~\cite{sarkar2015cost}.

\subsection{Progressive Sampling}
When the cost of collecting data is higher than the cost of building a performance model, it is imperative to minimize the number of measurements required for model building. A learning curve shows the relationship between the size of the training set and the accuracy of the model. In Figure~\ref{fig:learning_curve}, the horizontal axis represents the number of samples used to create the performance model, whereas the vertical axis represents the accuracy (measured in terms of MMRE) of the model learned. 
Learning curves typically have a steep sloping portion early in the curve followed by a plateau late in the curve. The plateau occurs when adding data does not improve the accuracy of the model. As engineers, we would like to stop sampling as soon as the learning curve starts to flatten.

Figure~\ref{fig:progressive_sampling} is a generic algorithm that defines the process of progressive sampling. \emph{Progressive} sampling starts by clearly defining the data used in the training set, called training pool, from which the samples would be selected (randomly, in this case) and then tested against the testing set. At each iteration, a (set of) data instance(s) of the training pool is added to the training set (Line 9). Once the data instances are selected from the training pool, they are evaluated, which in our setting means measuring the performance of the selected configuration (Line 12). The configurations and the associated performance scores are used to build  the model (Line 14). The model is validated using the testing set\footnote{The testing data consist of the configurations as well as the corresponding performance scores.}, then, the accuracy is then computed. The accuracy can be quantified by any measure, such as MMRE, MBRE, Absolute Error, etc. In our setting, we assume that the measure is accuracy (higher is better). Once the accuracy score is calculated, it is compared with the accuracy score obtained before adding the new set of configurations to the training set. If the accuracy of the model (with more data) does not improve the accuracy when compared to the previous iteration (lesser data), then a life is lost. This termination criterion is widely used in the field of multi-objective optimization to determine degree of convergence~\cite{krall2015gale}.

\begin{figure}[!t]
\small
\hspace{0.4cm}\begin{lstlisting}[xrightmargin=5.0ex, mathescape,frame=none,numbers=right]
  # Projective Sampling
  def projective(training, testing, thres_freq=3): 
    collector = list()
    independent_vals = list()
    dependent_vals = list()
    for count in range(1, len(training)):    
      # Add one configuration to the training set
      independent_vals += training[count]      
      # Measure the performance value for the newly
      # added configuration 
      dependent_vals += measure(training_set[count])  
      # update feature frequency table 
      T = update_frequency_table(training[count])
      # Build model
      model = build_model(independent_vals, dependent_vals)     
      # Test Model
      perf_score = test_model(model, testing, measure(testing))
      # Collect the the pair of |training set| 
      # and performance score
      collector += [count, perf_score]
      # minimum values of the feature frequency table
      if min(T) >= thresh_freq: break
    return model
\end{lstlisting}
\caption{\small{Pseudocode of projective sampling.}
}
\label{fig:projective_sampling} 
\end{figure}

\subsection{Projective Sampling}\label{sec:soa}

One of the shortcomings of progressive sampling is that the resulting 
performance model achieves an acceptable accuracy only after a large number of iterations, which implies  high modelling cost. There is no way to actually determine the cost of modelling until the performance model is already built, which defeats its purpose, as there is a risk of over-shooting the modelling budget and still not obtain an accurate model. \emph{Projective} sampling addresses this problem by approximating the learning curve using a minimal set of initial sampling points (configurations), thus providing the stakeholders with an estimate of the modelling cost. Sarkar et. al~\cite{sarkar2015cost} used projective sampling to predict the number of samples required to build a performance model. The initial data points are selected by randomly adding a constant number of samples (configurations) to the training set from the training pool. In each iteration, the model is built, and the accuracy of the model is calculated using the testing data. 
A feature-frequency heuristic is used as the termination criterion.
The feature-frequency heuristic counts the number of times a feature has been selected and deselected. Sampling stops when the counts of features selected and deselected is, at least, at a predefined threshold (\textit{thresh\_freq}).

Figure~\ref{fig:projective_sampling} provides a generic algorithm for projective sampling. Similar to progressive sampling, projective sampling starts with selecting samples from the training pool and adding them to the training set (Line 8). Once the samples are selected, the corresponding configurations are evaluated (Line 11). The feature-frequency table T is then updated by calculating the number of features that are selected and deselected in \textit{independent\_vals} (Line 13). The configurations and the associated performance values are then used to build a performance model, and the accuracy is calculated (Lines 15--17). The number of configurations and the accuracy score are stored in the collector, since our objective is to estimate the learning curve.  $min(T)$ holds the minimum value of the feature selection and deselection frequencies in T. Once the value of $min(T)$ is greater than  \textit{thresh\_freq}, the sampled points are used to estimate the learning curve. These points are used to search for a best-fit function that can be used to extrapolate the learning curve (there are several available, including Logarithmic, Weiss and Tian, Power Law and Exponential~\cite{sarkar2015cost}). Once the best-fit function is found, it is used to determine the point of convergence.

\section{Rank-based approach}

Typically, performance models are evaluated based on the accuracy or error. The error can be computed using\footnote{Aside: There has been a lot of criticism regarding MMRE, which shows that MMRE along with other accuracy statistics such as MMRE, MBRE has been shown to cause conclusion instability~\cite{myrtveit2012validity, myrtveit2005reliability, foss2003simulation}.}:
\begin{equation}\label{eq:err}
\mathit{MMRE}=\frac{\mid\mathit{predicted} - \mathit{actual}\mid}{\mathit{actual}} \cdot 100
\end{equation}

The key idea in this paper is to use ranking as an approach for building regression models. There are a number of advantages of using a rank-based approach:
\begin{itemize}
    \item For the use cases listed in the introduction, \textit{ranking is the ultimate goal}. A user may just want to identify the top-ranked configurations rather than to rank the whole space of configurations. For example, a practitioner trying to optimize an Apache Web server is searching for a set of configurations that can handle maximum load, and is not interested in the whole configuration space. 
    
    \item \textit{Ranking is extremely robust} since it is only mildly affected by errors or outliers~\cite{kloke2012rfit, rosset2005ranking}. Even though measures such as Mean Absolute Error are robust, in the configuration setting, a practitioner is often more interested in knowing the rank rather than the predicted performance scores.

    \item \textit{Ranking reduces the number of training samples required to train a model}. We will demonstrate that the number of training samples required to find the optimal configuration using a rank-based approach is reduced considerably, compared to residual-based approaches which use MMRE.
\end{itemize}

\begin{figure}[t]
\small
\hspace{0.4cm}\begin{lstlisting}[xrightmargin=5.0ex,mathescape,frame=none,numbers=right]
  # rank-based approach
  def rank_based(training, testing, lives=3): 
    last_score = -1
    independent_vals = list()
    dependent_vals = list()
    for count in range(1, len(training)):    
      # Add one configuration to the training set
      independent_vals += training[count]      
      # Measure the performance value for the newly
      # added configuration 
      dependent_vals += measure(training_set[count])
      # Build model
      model = build_model(independent_vals, dependent_vals)     
      # Predicted performance values
      predicted_performance = model(testing) 
      # Compare the ranks of the actual performance 
      # scores to ranks of predicted performance scores
      actual_ranks = ranks(measure(testing))
      predicted_ranks = ranks(predicted_performance)
      mean_RD = RD(actual_ranks, predicted_ranks)
      # If current rank difference is not better than
      # the previous rank difference, then loose life
      if mean_rank_difference <= last_rank_difference:
        lives -= 1
      last_rank_difference = mean_RD
      # If all lives are lost, exit loop
      if lives == 0: break
    return model

\end{lstlisting}
\caption{\small{Psuedocode of rank-based approach.}
}
\label{fig:rank-based}  
\end{figure}
It is important to note that we aim at building a performance model similar to the accurate performance model building process used by prior work as described in Section~\ref{sec:residual}. But instead of using residual measures of errors, as described in Equation~\ref{eq:err}, which depend on residuals ($r = y - f(x)$), \footnote{Refer to Section~\ref{sec:problem_formal} for definitions.} we use a rank-based measure. While training the performance model ($f(x)$), the configuration space is iteratively sampled (from the training pool) to train the performance model. Once the model is trained, the accuracy of the model is measured  by sorting the values of $y=f(x)$ from `small' to `large', that is:
\begin{equation}
    f(x_1) \le f(x_2) \le f(x_3) \le ... \le f(x_n).
\end{equation}
The predicted rank order is then compared to the actual rank order. The accuracy is calculated using the mean rank difference:
\begin{equation} \label{eq:rank_performance}
    \mathit{accuracy} = \frac{1}{n} \cdot \mathlarger{\mathlarger{\sum}}_{i=1}^{n}\abs[\Big]{rank(y_i) - rank(f(x_i))}
\end{equation}
This measure simply counts how many of the pairs in the test data were ordered incorrectly by the performance model $f(x)$ and measures the average of magnitude of the ranking difference.

In Figure~\ref{fig:rank-based}, we list a generic algorithm for our rank-based approach. Sampling starts by selecting samples randomly from the training pool and by adding them to the training set (Line 8). The collected sample configurations are then evaluated (Line 11). The configurations and the associated performance measure are used to build a performance model (Line 13). The generated model (CART, in our case) is used to predict the performance measure of the configurations in the testing pool (Line 16). Since the performance value of the testing pool is already measured, hence known, the ranks of the actual performance measures, and predicted performance measure are calculated. (Lines 18--19). The actual and predicted performance measure is then used to calculate the rank difference using Equation~\ref{eq:rank_performance}. If the rank difference of the model (with more data) does not decrease  when compared to the previous generation (lesser data), then a life is lost (Lines 23--24). When all lives are expired, sampling terminates (Line 27).

The motivation behind using the parameter \textit{lives} is: to detect convergence of the model building process. If adding more data does not improve the accuracy of the model (for example, in Figure~\ref{fig:learning_curve} the accuracy of the model generated does not improve after 20 samples configuration), the sampling process should terminate to avoid resource wastage; see also Section~\ref{sec:tradeoff}.

\section{Subject Systems}\label{sec:systems}
To compare residual-based approaches with our rank-based approach, we evaluate it using 21 test cases collected in 9 open-source software systems\footnote{For more details on the subject systems and configurations options refer to \url{http://tiny.cc/3wpwly}}. 
\begin{enumerate}
 \item \hlgreen{SS1} \textsc{x264} is a video-encoding library that encodes video streams to H.264/MPEG-4 AVC format. We consider 16 features, which results in 1152 valid configurations.
 
 \item \hlgreen{SS2} \textsc{Berkeley DB (C)} is an embedded key-value-based database library that provides scalable high performance database management services to applications. We consider 18 features resulting in 2560 valid configurations.
 
 \item \hlgreen{SS3} \textsc{SQLite} is the most popular lightweight relational database management system. It is used by several browsers and operating systems as an embedded database. In our experiments, we consider 39 features that give rise to more than 3 million valid configurations.
 
 \item \hlgreen{SS4} \textsc{wget} is a software package for retrieving files using HTTP, HTTPS, and FTP. It is a non-interactive command line tool. In our experiments, we consider 16 features, which result in 188 valid configurations.

\item \hlyellow{SS5} \textsc{lrzip} or Long Range ZIP is a compression program optimized for large files, consisting mainly of an extended rzip step for long distance redundant reduction and a normal compressor step. We consider 19 features, which results in 432 valid configurations.

\item \hlyellow{SS6} \textsc{Dune} or the Distributed and Unified Numerics Environment, is a modular C++ library for solving partial differential equations using grid-based methods. We consider 11 feature resulting in 2305 valid configurations.

\item \hlyellow{SS7} \textsc{HSMGP} or Highly Scalable MG Prototype is a prototype code for benchmarking Hierarchical Hybrid Grids data structures, algorithms, and concepts. It was designed to run on super computers. We consider 14 features resulting in 3456 valid configurations.

\item \hlyellow{SS8} \textsc{Apache} HTTP Server is a Web Server; we consider 9 features resulting in 192 valid configurations.

\end{enumerate}

In addition to these 8 subject systems, we also consider Apache Storm, a distributed system, in several scenarios. The datasets were obtained from the paper by Jamshidi et al.~\cite{jamshidi2016uncertainty}. The experiment considers three  benchmarks namely:
\begin{itemize}
    \item WordCount (\textsc{wc}) counts the number of occurences of the words in a text file. 
    \item RollingSort (\textsc{rs}) implements a common pattern in real-time analysis that performs rolling counts of messages. 
    \item SOL (\textsc{sol}) is a network intensive topology, where the message is routed through an inter-worker network.
\end{itemize}
The experiments were conducted with all the above mentioned benchmarks on 5 cloud clusters. The experiments also contain measurement variabilities, the \textsc{wc} experiments were also carried out on multi-tenant cluster, which were shared with other jobs. For example, \textsc{wc}+\textsc{rs} means \textsc{wc} was deployed in a multi-tenant cluster with \textsc{rs} running on the same cluster. As a result, not only latency increased but also variability became greater. The environments considered in our experiments are:
\begin{enumerate}
    \setcounter{enumi}{8}
    \item \hlyellow{SS9} \textsc{\textsc{wc}-6d-throughput} is an environment configuration where \textsc{wc} is executed by varying 6 features resulting in 2879 configurations;  throughput is calculated.

    \item \hlyellow{SS10} \textsc{\textsc{rs}-6d-throughput} is an environment configuration where \textsc{rs} is run by varying 6 features which results in 3839 configurations; the throughput is measured.
    
    \item \hlyellow{SS11} \textsc{\textsc{wc}-6d-latency} is an environment configuration where \textsc{wc} is executed by varying 6 features resulting in 2879 configurations; latency is calculated.
    
    \item \hlyellow{SS12} \textsc{\textsc{rs}-6d-latency} is an environment configuration where \textsc{rs} is executed by varying 6 features, which results in 3839 configurations; latency is measured.
    
    \item \hlyellow{SS13} \textsc{ \textsc{wc}+\textsc{rs}-3d-throughput} is an environment configuration where \textsc{wc} is run in a multi-tenant cluster along with \textsc{rs}. \textsc{wc} is executed by varying 3 features resulting in 195 configurations; throughput is measured.
    
    \item \hlyellow{SS14} \textsc{\textsc{wc}+\textsc{sol}-3d-throughput} is an environment configuration where \textsc{wc} is run in a multi-tenant cluster along with \textsc{sol}. \textsc{wc} is executed by varying 3 features resulting in 195 configurations; throughput is measured.
    
    \item \hlred{SS15} \textsc{\textsc{wc}+\textsc{wc}-3d-throughput} is an environment configuration where \textsc{wc} is run in a multi-tenant cluster along with \textsc{wc}. \textsc{wc} is executed by varying 3 features resulting in 195 configurations; throughput is measured.
    
    \item \hlred{SS16} \textsc{\textsc{sol}-6d-throughput} is an environment configuration where \textsc{sol} is executed by varying 6 features resulting in 2865 configurations; throughput is measured.
        
    \item \hlred{SS17} \textsc{\textsc{wc}-\textsc{wc}-3d-throughput} is an environment configuration where \textsc{wc} is executed by varying 3 features resulting in 755 configurations; throughput is calculated.
    
    \item \hlred{SS18} \textsc{\textsc{wc}+\textsc{sol}-3d-latency} is an environment configuration where \textsc{wc} is run in a multi-tenant cluster along with \textsc{sol}. The \textsc{wc} is executed by varying 3 features resulting in 195 configurations; latency is measured.
    
    \item \hlred{SS19} \textsc{\textsc{wc}+\textsc{wc}-3d-latency} is an environment configuration where \textsc{wc} is run in a multi-tenant cluster along with \textsc{wc}. The \textsc{wc} is executed by varying 3 features resulting in 195 configurations; latency is measured.

    \item \hlred{SS20} \textsc{\textsc{sol}-6d-latency} is an environment configuration where \textsc{sol} is executed by varying 6 features resulting in 2861 configuration setting; latency is measured.
    
    \item \hlred{SS21} \textsc{\textsc{wc}+\textsc{rs}-3d-latency} is an environment configuration where \textsc{wc} is run in a multi-tenant cluster along with \textsc{rs}. \textsc{wc} is executed by varying 3 features resulting in 195 configurations; latency is measured.

\end{enumerate}

\section{Evaluation}
\subsection{Research Questions}
In the past, configuration ranking required an accurate model of the configuration space, since an inaccurate model implicitly indicates that the model has missed the trends of the configuration space. Such accurate models require the evaluation/measurement of hundreds of configuration options for training~\cite{siegmund2012predicting, guo2013variability, sarkar2015cost}. There are also cases where building an accurate model is not possible, as shown in Figure~\ref{fig:model_efficiency} (right side). 

Our research questions are geared towards finding optimal configurations when building an accurate model of a given software system is not possible. As our approach relies on ranking, our hypothesis is that we would be able to find the (near) optimal configuration  using our rank-based approach while using fewer measurements, as compared to an accurate model learnt using  residual-based approaches\footnote{Aside: It is worth keeping in mind that the approximation error in a model does not always harm. A model capable to smoothing the complex landscape of a problem can be beneficial for the search process. This sentiment has been echoed in the evolutionary algorithm literature as well~\cite{lim2010generalizing}.}.

Our proposal is to embrace rank preservation but with inaccurate models and to use these models to guide configuration rankings. Therefore, to assess the feasibility and usefulness of the inaccurate model in configuration rankings, we consider the following:
\begin{itemize}
    \item Accurate rankings found by inaccurate models using a rank-based approach, and
    \item the effort (number of measurements) required to build an inaccurate model. 
\end{itemize}
The above considerations lead to  two research questions:

\textbf{RQ1}\textit{: Can  inaccurate    models accurately rank
configurations?}
Here, the optimal configurations found using an inaccurate model are compared to the more accurate models generated using residual-based approaches. The accuracy of the models is calculated using MMRE (from Equation~\ref{eq:err}). 


\textbf{RQ2}\textit{: How expensive is a rank-based approach (in terms of
 how many configurations must be executed)?}
It is expensive to build accurate models, and our goal is to minimize the number of measurements. It is important to demonstrate that we can find optimal configurations of a system using inaccurate models as well as reducing the number of measurements.

\begin{figure*}[tbh]
\centering
\includegraphics[width=0.8\paperwidth, height=4.7cm]{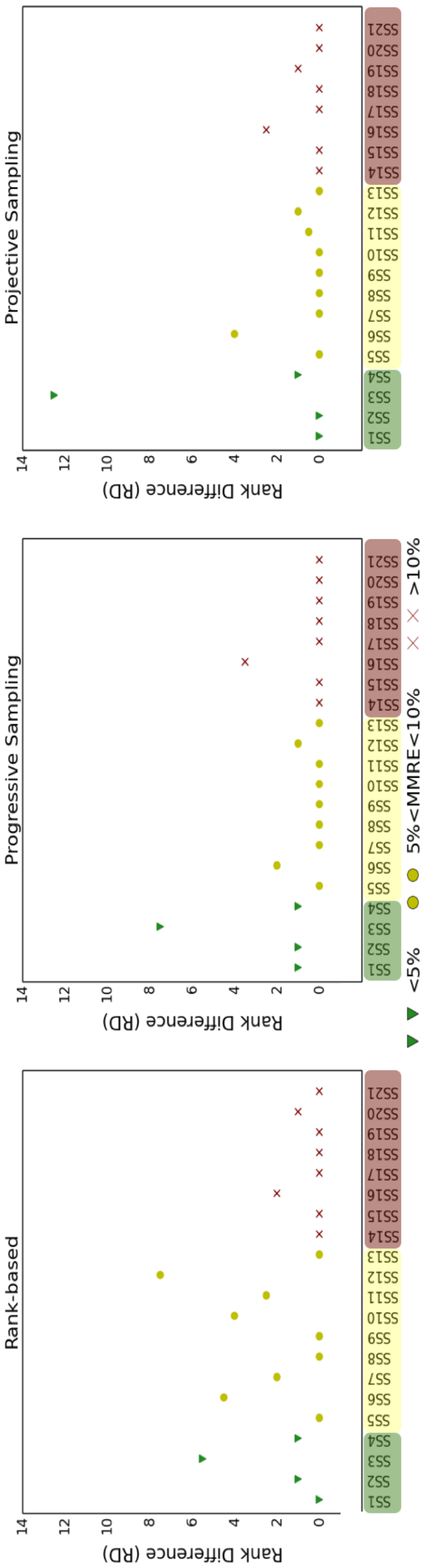}
\caption{
    \small{The rank difference of the prediction made by the model built using residual-based and rank-based approaches. Note that the y-axis of this chart rises to some very large values; e.g., SS3 has over three million
    possible configurations. Hence, the above charts could be summarised as follows: ``the rank-based approach is surprisingly accurate since the rank difference is usually close to 0\% of the total number of possible configurations''. In this figure, \protect\custriangle{}, \protect\cusdot{}, and \protect\cuscross{} represent the subject systems (using the technique mentioned at the top of the figure), in which we could build a prediction model, where accuracy is  $< 5\%$, $5\%<MMRE<10\%$, and $>10\%$ respectively. This is based on Figure~\ref{fig:model_efficiency}.
    }
}\label{fig:rankdiff}
\end{figure*}

\subsection{Experimental Rig}
For each subject system, we build a table of data, one row per valid configuration. We then run all configurations of all systems and record the performance scores (i.e., that are invoked by a benchmark). The exception is SQLite, for which we measure only 4400 configurations corresponding to feature-wise and pair-wise sampling and additionally 100 random configurations. (This is because SQLite has 3,932,160 possible configurations,
which is an impractically large number of configurations to test.) To this table, we added a column showing the performance score obtained from the actual measurements for each configuration. Note that, while answering the research questions, we ensure that we never test any prediction model on the data that we used to learn the model. Next, we repeat the following procedure 20 times. 

To answer the research questions, we split the datasets into training pool (40\%), testing pool (20\%), and validation pool (40\%). The experiment is conducted in the following way:
\begin{itemize}
    \item Randomize the order of rows  in  the training data
    \item \textbf{Do}
    \begin{itemize}
        \item Select one configuration (by sampling with replacement) and add it to the training set
        \item Determine the performance scores associated with the configuration. This corresponds to  a table look-up, but would entail compiling or configuring and executing a system configuration in a practical setting.
        \item Using the training set and the accuracy, build a performance model using CART.
        \item Using the data from the testing pool, assess the accuracy either using MMRE (as described in Equation~\ref{eq:err}) or the rank difference (as described in Equation~\ref{eq:rank_performance}).         
    \end{itemize}
    \item \textbf{While} the accuracy is greater or equal to the threshold determined by the practitioner (rank difference in the case of our rank-based approach and MMRE in the case of the residual-based approaches).
\end{itemize}

Once the model has been iteratively trained, it is used on the data in the validation pool. Please note, the learner has not been trained on the validation pool.
RQ1 relates the results found by the inaccurate performance models (rank-based) to more accurate models (residual-based). We use the absolute difference between the ranks of the configurations predicted to be the optimal configuration and the actual optimal configuration. We call this measure rank difference (\textit{RD}).  
\begin{equation}\label{eq:rankdiff}
    \begin{split}
        RD &= \abs[Big]{\mathit{rank(actual_{optimal})} - \mathit{rank(predicted_{optimal})}}\\
    \end{split}
\end{equation}

\noindent Ranks are calculated by sorting the configurations based on their performance scores. The configuration with the least performance score, $\mathit{rank(actual_{optimal})}$, is ranked 1 and the one with highest  score is ranked as $N$, where $N$ is the number of configurations.




\section{Results}
\subsection{RQ1: \textit{Can  inaccurate    models accurately rank configurations?}}

\begin{figure}[tbh]
    {
        {\scriptsize \begin{tabular}{|l@{~~~}|l@{~~~}|r@{~~~}|r@{~~~}|c|}
            \hline
            \textbf{Rank} & \textbf{Treatment} & \textbf{Median} & \textbf{IQR} & \textbf{Median and IQR chart}\\\hline
            
            \rowcolor{lightgray}
            \textbf{SS1}  & \textbf{} & \textbf{} & \textbf{}& \\\hline            
                1 & Projective &    0.0  &  0.0 & \quart{0}{0}{0} \\
                  1 & Progressive &    0.0  &  1.0 & \quart{0}{9}{0} \\
                2 &  Rank-based &    2.0  &  8.0 & \quart{0}{79}{19} \\

            \hline \rowcolor{lightgray}
            \textbf{SS2}  & \textbf{} & \textbf{} & \textbf{}& \\\hline
            
              1 & Projective &    0.0  &  1.0 & \quart{0}{4}{0} \\
            2 &  Rank-based &    1.0  &  6.0 & \quart{0}{25}{4} \\
              2 & Progressive &    2.0  &  18.0 & \quart{4}{75}{8} \\
              \rowcolor{lightgray}
            \textbf{SS3}  & \textbf{} & \textbf{} & \textbf{}& \\\hline
            
                1 & Progressive &    10.0  &  17.0 & \quart{0}{9}{5} \\
                  2 & Projective &    15.0  &  139.0 & \quart{0}{79}{8} \\
                  2 &  Rank-based &    21.0  &  40.0 & \quart{1}{23}{11} \\

            \hline \rowcolor{lightgray}
            \textbf{SS11}  & \textbf{} & \textbf{} & \textbf{}& \\\hline  
                
                  1 & Progressive &    0.0  &  1.0 & \quart{0}{26}{0} \\
                2 &  Rank-based &    1.0  &  3.0 & \quart{0}{79}{26} \\
                  2 & Projective &    1.0  &  2.0 & \quart{0}{53}{26} \\

                \hline \rowcolor{lightgray}
            \textbf{SS20}  & \textbf{} & \textbf{} & \textbf{}& \\\hline   
                  1 & Projective &    0.0  &  1.0 & \quart{0}{3}{0} \\
  1 & Progressive &    0.0  &  1.0 & \quart{0}{3}{0} \\
 2 &  Rank-based &    5.0  &  19.0 & \quart{3}{76}{19} \\
                              
            \hline  \end{tabular}}
    }
    
    \caption{
        {\small 
            Median rank difference of 20 repeats. Median ranks is the rank difference  as described in Equation~\ref{eq:rankdiff}, and IQR the difference between 75th percentile and 25th percentile found during multiple repeats. 
            Lines with a dot in the middle 
            (~\protect\quartex{-2}{13}{6}), 
            show the median as a round dot within the IQR.
            All the results are sorted by the median rank difference: a lower median value is better. 
            The left-hand column (\textit{Rank}) ranks the various techniques for example, when comparing various techniques for SS1, a rank-based approach has a different rank since their median rank difference is statistically different. 
        }
    }
    \label{fig:stat-test}
\end{figure}

\begin{figure*}[t]
        \centering
        \includegraphics[width=0.7\paperwidth, height=5.4cm]{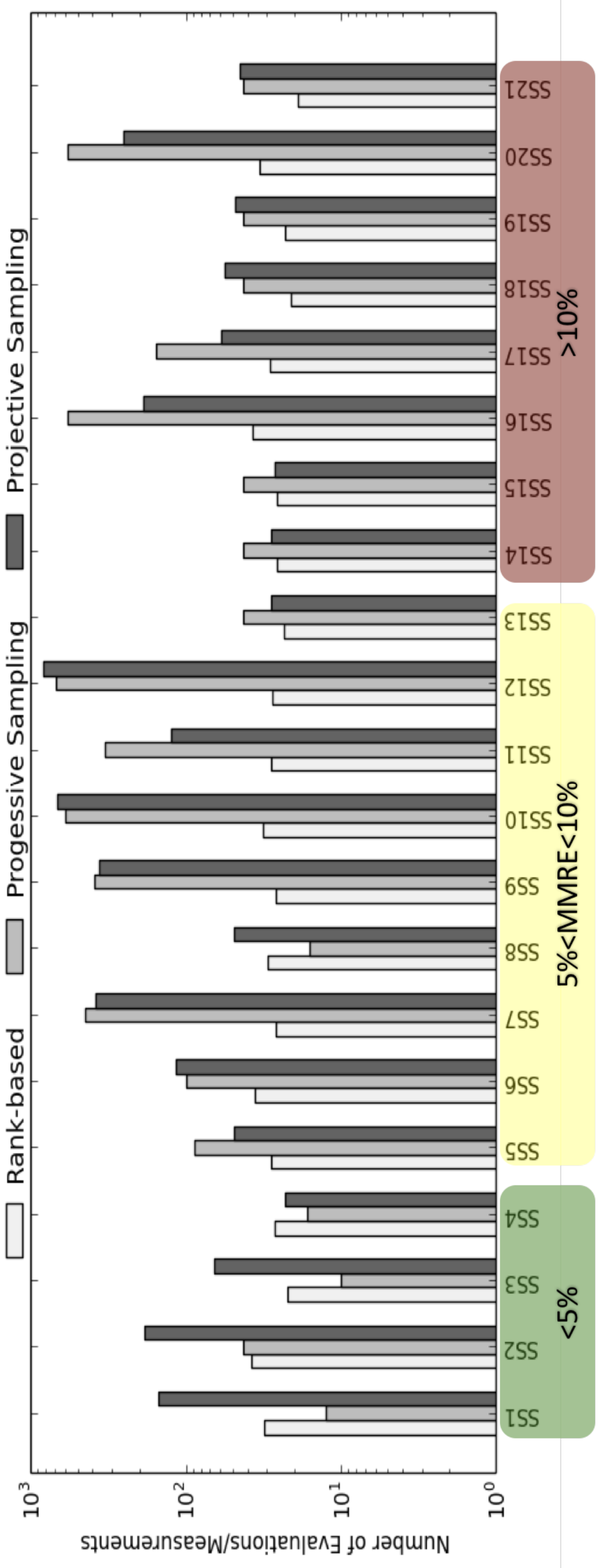}

        \caption{
        \small{
        Number of measurements required to train models  by different approaches. The software systems are ordered based on the accuracy scores of Figure~\ref{fig:model_efficiency}. }
        }\label{fig:evals}
\end{figure*}

Figure~\ref{fig:rankdiff} shows the \textit{RD} of the predictions built using the rank-based approach and residual-based approaches \footnote{The MMRE scores for the models \url{http://tiny.cc/bu14iy}} by learning from 40\% of the training set and iteratively adding data to the training set (from the training pool), while testing against the testing set (20\%). The model is then used to find the optimal configuration among the configurations in the validation dataset (40\%). The horizontal axis shows subject systems. The vertical axis shows the rank difference (\textit{RD}) from Equation~\ref{eq:rankdiff}. In this figure:
\begin{itemize}
    \item The perfect performance model would be able to find the optimal configuration. Hence, the ideal result of this figure would be if all the points lie on the $y=0$ or the horizontal axis. That is, the model was able to find the optimal configuration for all the subject systems (\textit{RD} = 0). 
    \item The markers \protect\custriangle{}, \protect\cusdot{}, and \protect\cuscross{} represent the software systems where  a  model with a certain accuracy can be built, measured in MMRE is  $<5\%$, $5\%<$MMRE$<10\%$, and $>10\%$ respectively.
\end{itemize}

Overall, in Figure~\ref{fig:rankdiff}, we find that:
\begin{itemize}
    \item The \cuscross{} represents software systems where the performance models are inaccurate ($>10\%$ MMRE) and still can be used for ranking configurations, since the rank difference of these systems is always less than 4. Hence, even an inaccurate performance model can rank configurations.
    \item All three models built using both rank-based and residual-based approaches are able to find near optimal configurations. For example, progressive sampling for SQLite predicted the configuration whose performance score is ranked 9th in the testing set. This is good enough since progressive sampling is able to find the 9th most performant configuration among 1861 configurations\footnote{Since we test only on 40\% of the possible configuration space (40\% of 4653).}.
    \item The mean rank difference of the $\mathit{predicted}_{\mathit{\small{optimal}}}$ is 1.4, 0.77, and 0.93\footnote{The median rank difference is 0 for all the approaches.} for the rank-based approach, progressive sampling, and projective sampling respectively. Thus, a performance model can be used to rank configurations.
\end{itemize}

We claim that the rank of the optimal configuration found by the residual and rank-based approaches is the same. To verify that the similarity is statistically significant, we further studied the results using non-parametric tests, which were used by Arcuri and Briand at ICSE'11~\cite{mittas13}. For testing statistical significance,
we used a non-parametric bootstrap test with 95\% confidence~\cite{efron93}, followed by
an A12 test to check that any observed differences were not trivially small effects;
that is, given two lists $X$ and $Y$, count how often there are larger
numbers in the former list (and if there are ties, add a half mark):
$a=\forall x\in X, y\in Y\frac{\#(x>y) + 0.5\cdot\#(x=y)}{|X|\cdot|Y|}$
(as per Vargha~\cite{Vargha00}, we say that a ``small'' effect has $a <0.6$). 
Lastly, to generate succinct reports, we use the Scott-Knott test to recursively
divide our approaches. This recursion used A12 and bootstrapping  
to group together subsets that are (a)~not significantly different and are (b)~not
just a small effect different to each other. This use of the Scott-Knott test is endorsed
by Mittas and Angelis~\cite{mittas13}
and by Hassan et al.~\cite{7194626}.


In Figure~\ref{fig:stat-test}, the table shows the Scott-Knott ranks for the three approaches. The quartile charts are the Scott-Knott results for our subject systems, where the rank-based approach did not do as well as the residual-based approaches~\footnote{For complete Skott-Knott charts, refer to \url{http://geekpic.net/pm-1GUTPZ.html}.}. For example, the statistic test for SS1 shows that the ranks of the optimal configuration by the rank-based approach was statistically different from the ones found by the residual-based approaches. We think this is reasonably close since the median rank  found by the rank-based approach is 2 out of 460 configurations, whereas residual-based approaches find the optimal configurations with a median rank of 0. As our motivation was to find optimal configurations for software systems for which performance models were difficult or infeasible to build, we look at SS20. If we look at the Skott-Knott chart for SS20, the median rank found by the rank-based approach is 5, whereas the residual-based approaches could find the optimal configurations very consistently (IQR=1). But as engineers, we feel that this is close because we are able to find the 5th best configuration using 33 measurements compared to 251 and 576 measurements used for progressive and projective sampling, respectively. Overall, our results indicate that:
\vskip 1ex
 \begin{myshadowbox}
         A rank preserving (probably inaccurate) model can be useful in finding (near) optimal configurations of a software system using a rank-based approach.
 \end{myshadowbox}

\subsection{RQ2: {\em How expensive is a rank-based approach?}}
\begin{figure*}[t]
        \centering
        \includegraphics[width=0.7\paperwidth, height=6cm]{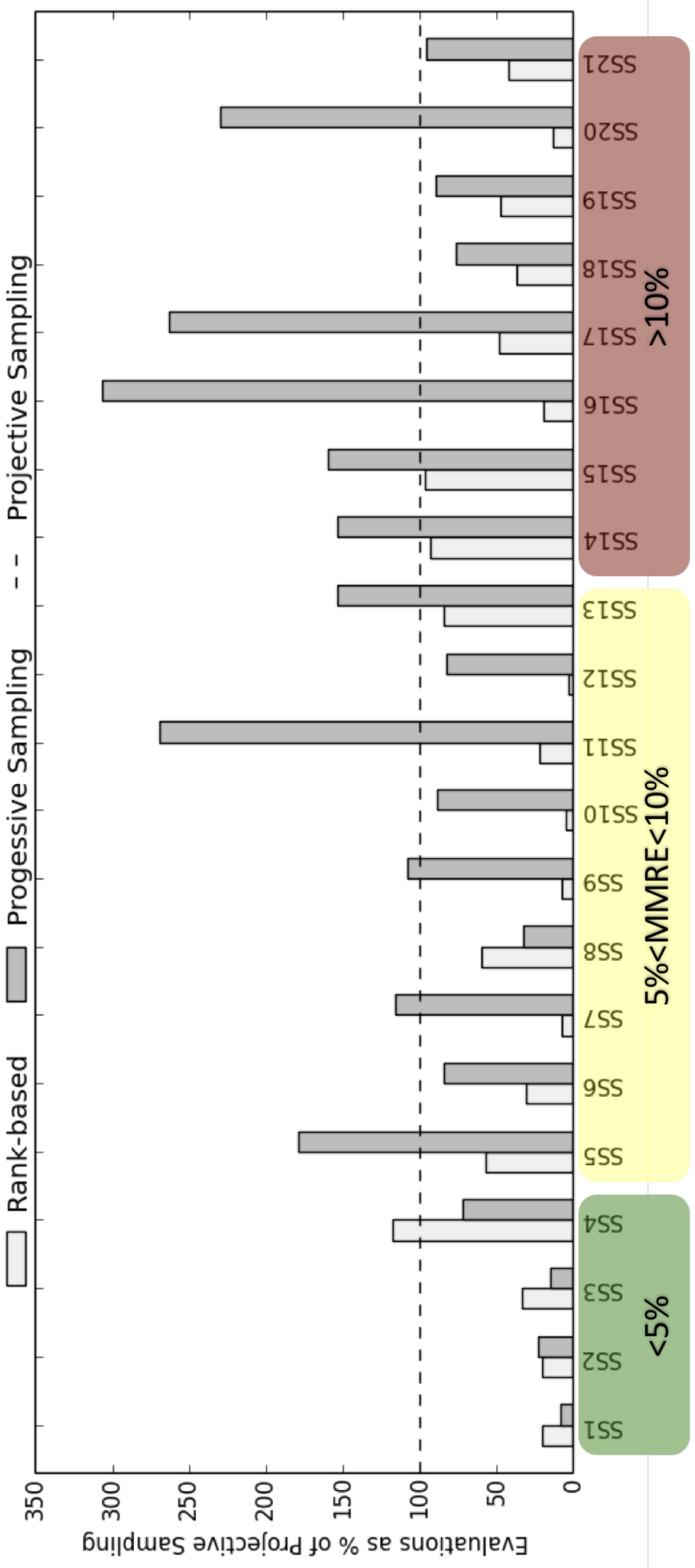}
        \caption{
        \small{The percentage of measurement used for training models with respect to the number of measurements used by projective sampling (dashed line). The rank-based approach uses almost 10 times less measurements than the residual-based approaches. The subject systems are ordered based on the accuracy scores of Figure~\ref{fig:model_efficiency}.}
        }\label{fig:evals_ratio}
\end{figure*}

To answer the question of whether we can recommend the rank-based approach as a cheap method for finding the optimal configuration, it is important to demonstrate that rank-based models are indeed cheap to build. In our setting, the cost of a model is determined by the number of measurements required to train the model. Figure~\ref{fig:evals} demonstrates this relationship. The vertical axis denotes the number of measurements in log scale and horizontal axis represents the subject systems.

In the systems SS1--SS4 (green band), the number of measurements required by the rank-based approach is less than for projective sampling and more than for progressive sampling. This is because the subject systems are easy to model.
For the systems SS4--SS13 (yellow band), the number of measurements required to build models using the rank-based approach is less than residual-based approaches, with the exception of SS8. Note that, as building accurate models becomes difficult,  the difference between the number of measurements required by the rank-based approach and residual-based approaches increases. 
For the systems SS14--SS21 (red band), the number of measurements required by the rank-based approach to build a model is always less than for residual-based approaches, with significant gains for SS19--SS21. 

In Figure~\ref{fig:evals_ratio}, the ratio of the measurements of different approaches are represented as the percentage of number of measurements required by projective sampling -- since it uses the most measurements in 50\% of the subject systems. For example, in SS5, the number of measurements used by progressive sampling is twice as much as used by projective sampling, whereas the rank-based approach uses half of the total number of measurements used by projective sampling. We observe that the number of measurements required by the rank-based approach is much lower than for the residual-based approaches, with the only exceptions of SS4 and SS8. 
We argue that such outliers are not of a big concern since the motivation of rank-based approach is to find optimal configurations for software systems, where an accurate model is  infeasible.

To summarize, the number of samples required by the rank-based approach is much smaller than for residual-based approaches. There are $\frac{4}{21}$ cases where residual-based approaches (progressive sampling) use fewer measurements. The subject systems where residual-based approaches use fewer measurements are systems where accurate models are easier to build (green and yellow band):
\vskip 1ex
\begin{myshadowbox}
 Models built using the rank-based approach require fewer measurements than residual-based approaches. In $\frac{8}{21}$ of the cases, the number of measurements is an order of magnitude smaller than residual-based approaches.
\end{myshadowbox}

\begin{figure}[t]
\centering
\includegraphics[scale=0.2]{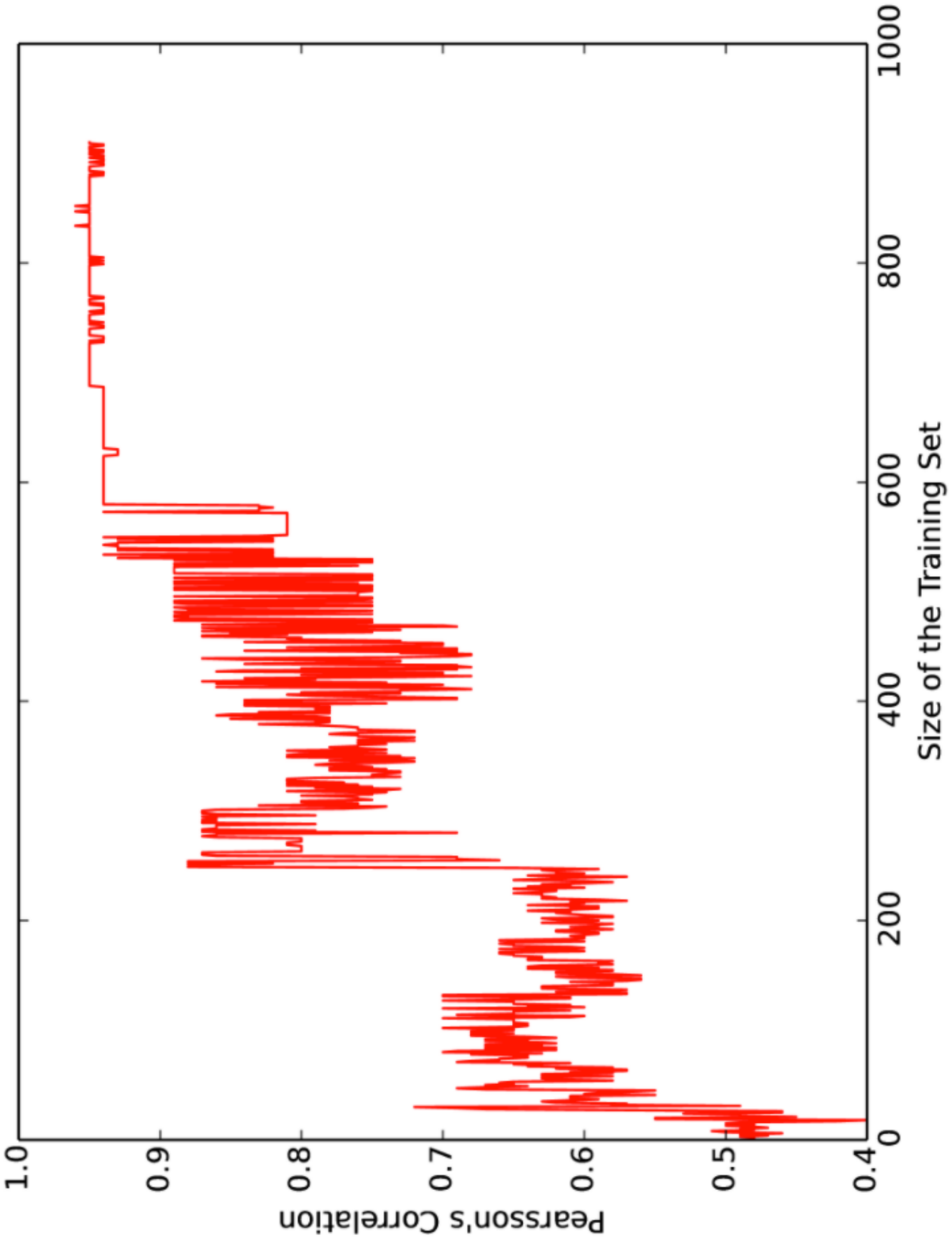}
\caption{\small The correlation between actual and predicted performance values (not ranks) increases as the training set increases. This is evidence that the model does learn as training progresses. 
}\label{fig:correl}
\end{figure}

\section{Discussion}

\subsection{How is rank difference useful in configuration optimization?}

The objective of the modelling task considered here is to assist practitioners to find optimal configuration/s. We use a performance model, like traditional approaches, to solve this problem, but rather than using a residual-based accuracy measure, we use rank difference. Rank difference can also be thought as a correlation-based metric, since rank difference essentially tries to preserve the ordering of performance scores. 

In our rank-based approach, we do not train the model to predict the dependent values of the testing set. But rather, we attempt to train the model to predict the dependent value that is correlated to the actual values of the testing set. So, during iterative sampling based on the rank-based approach, we should see an increase in the correlation coefficient as the training progresses. Figure~\ref{fig:correl} shows how the correlation between actual and the predicted values increases as the size of the training set grows. From the combination of Figure~\ref{fig:model_efficiency} and Figure~\ref{fig:rankdiff}, we see that even an inaccurate model can be used to find an optimal configuration.\footnote{This also shows how a correlation-based measure can be used as a stopping criterion.}

\subsection{Can inaccurate models be built using residual-based approaches?}
We have already shown that a rank preserving (probably inaccurate) model is sufficient to find the optimal configuration of a given system. MMRE can be used as a stopping criterion, but as we have seen with residual-based approaches, they require a larger training set and hence are not cost effective. This is because, with residual-based approaches, unlike our rank-based approach, it is not possible to know when to terminate sampling. 
It may be noted that rank difference can be easily replaced with a correlation-based approach such as Pearson's or Spearman's correlation.

\subsection{Can we predict the complexity of a system to determine which approach to use?}
From our results, we observe that a rank-based approach is not as effective as the residual-based approaches for software systems that can be modelled accurately (\hlgreen{green} band). Hence, it is important to distinguish between software systems where the rank-based approach is suitable and software systems where residual-based approaches are suitable. This is relatively straight-forward since both rank-based and residual-based approaches use random sampling to select the samples. The primary difference between the approaches is the termination criterion. The rank-based approach uses rank difference as the termination criterion whereas residual-based approaches use criterion based on MMRE, etc. Hence, it is possible to use both techniques simultaneously. If a practitioner observes that the accuracy of the model during the building process is high (as in case of SS2), residual-based approaches would be preferred. Conversely, if the accuracy of the model is low (as in the case of 
, SS21), the rank-based approach would be preferred.

\begin{figure}[t]
\centering
\includegraphics[scale=0.27]{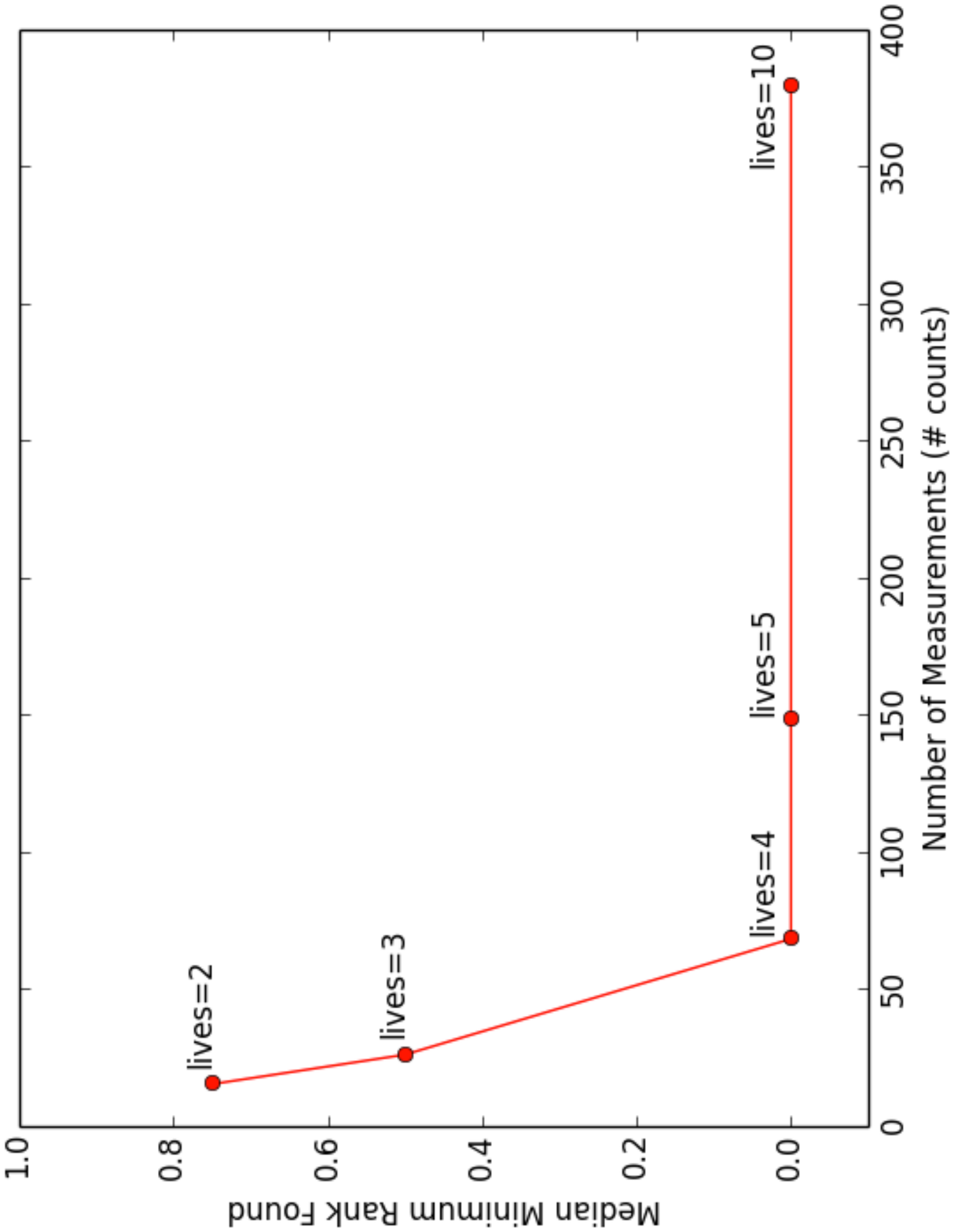}
\caption{\small The trade-off between the number of measurements or size of the training set and the number of \textit{lives}.
}\label{fig:paramT}
\end{figure}

\subsection{What is the trade-off between the number of lives and the number of measurements?}\label{sec:tradeoff}
Our rank-based approach requires that the practitioner defines a termination criterion (\textit{lives} in our setting) before the sampling process commences, which is a similar to progressive sampling. The termination criterion preempts the process of model building based on an accuracy measure. The rank-based approach uses rank difference as the termination criterion, whereas residual-based approaches use residual-based measures. In our experiments, the number of measurements or the size of the training set depends on the termination criterion~(\textit{lives}). An early termination of the sampling process would lead to a sub-optimal configuration, while late termination would result in resource wastage. Hence, it is important to discuss the trade-off between the number of \textit{lives} and the number of measurements.
In Figure~\ref{fig:paramT}, we show the trade-off between the median minimum ranks found and the number of measurements (size of training set). The markers of the figure are annotated with the values of \textit{lives}. The trade-off characterizes the relationship between two conflicting objectives, for example, point (\textit{lives}=2) requires very few measurements but the minimum rank found is the highest, whereas point (\textit{lives}=10) requires a large number of measurements but is able to find the best performing configuration. Note, this curve is an aggregate of the trade-off curves of all the software systems discussed in this paper\footnote{Complete trade-off curves can be found at \url{http://tiny.cc/kgs2iy} or \url{http://tiny.cc/rank-param}.}.  Since our objective is to minimize the number of measurements while reducing rank difference, we assign the value of $3$ to \textit{lives} for the purposes of our experiments.

\section{Reliability and Validity} 
{\em Reliability} refers to the consistency of the results obtained
from the research.  For example,   how well can independent researchers reproduce the study? To increase external
reliability, we took care to either  clearly define our
algorithms or use implementations from the public domain
(scikit-learn)~\cite{scikit-learn}. Also, all data and code used in this work are available
on-line.\footnote{ 
\url{https://github.com/ai-se/Reimplement/tree/cleaned_version}}

{\em Validity} refers to the extent to which a piece of research actually
investigates what the researcher purports to investigate~\cite{SSA15}.
{\em Internal validity} is concerned with whether the differences found in
the treatments can be ascribed to the treatments under study. 

For SQLite, we cannot measure all possible configurations in reasonable time. Hence, we sampled only 100 configurations to compare prediction and actual values. We are aware that this evaluation leaves room for outliers and that measurement bias can cause false interpretations~\cite{me12d}. Since we limit our attention to predicting performance for a given workload, we did not vary benchmarks.

We aimed at increasing {\em external validity} by choosing subject systems from different domains with different configuration mechanisms. Furthermore, our subject systems  are deployed and used in the real world.

\section{Conclusion}
Configurable systems are widely used in practice, but it requires careful tuning to adapt them to a particular setting. State-of-the-art approaches use a residual-based technique to guide the search for optimal configurations. The model-building process involves iterative sampling used along with a residual-based accuracy measure to determine the termination point. These approaches require too many measurements and hence are expensive. To overcome the requirement of a highly accurate model, we propose a rank-based approach, which requires a lesser number of measurements and finds the optimal configuration just by using the ranks of configurations as an evaluation criterion.

Our key findings are the following. First, a highly accurate model is not required for  configuration optimization of a software system. We demonstrated how a rank-preserving (possibly even inaccurate) model can still be useful for ranking configurations, whereas a model with accuracy as low as 26\% can be useful for configuration ranking. Second, we introduce a new rank-based approach that can be used to decide when to stop iterative sampling. We show how a rank-based approach is not trained to predict the raw performance score but rather learns the model, so that the predicted values are correlated to actual performance scores.

To compare the rank-based approach to the state-of-the-art residual-based approaches (projective and progressive sampling), we conducted  a number of experiments on 9 real-world configurable systems to demonstrate the effectiveness of our approach. We observed that the rank-based approach is effective to find the optimal configurations for most subject systems while using fewer measurements than residual-based approaches. The only exceptions are subject systems for which building an accurate model is easy, anyway.
 
\section*{Acknowledgement}
The work is partially funded by an NSF award \#1302169. Siegmund\textquotesingle s work is supported by the DFG under the contract SI 2171/2. Apel\textquotesingle s work is supported by the DFG under the contract AP 206/6 and AP 206/7.


%

\balance
\bibliographystyle{ACM-Reference-Format}

\input{main.bbl}

\end{document}

%% file: main.bbl

%% file: main.bbl

\begin{thebibliography}{00}


\ifx \showCODEN    \undefined \def \showCODEN     #1{\unskip}     \fi
\ifx \showDOI      \undefined \def \showDOI       #1{{\tt DOI:}\penalty0{#1}\ }
  \fi
\ifx \showISBNx    \undefined \def \showISBNx     #1{\unskip}     \fi
\ifx \showISBNxiii \undefined \def \showISBNxiii  #1{\unskip}     \fi
\ifx \showISSN     \undefined \def \showISSN      #1{\unskip}     \fi
\ifx \showLCCN     \undefined \def \showLCCN      #1{\unskip}     \fi
\ifx \shownote     \undefined \def \shownote      #1{#1}          \fi
\ifx \showarticletitle \undefined \def \showarticletitle #1{#1}   \fi
\ifx \showURL      \undefined \def \showURL       #1{#1}          \fi
\providecommand\bibfield[2]{#2}
\providecommand\bibinfo[2]{#2}
\providecommand\natexlab[1]{#1}
\providecommand\showeprint[2][]{arXiv:#2}

\bibitem[\protect\citeauthoryear{Chen, Nair, Krishna, and Menzies}{Chen
  et~al\mbox{.}}{2016}]%
        {chen2016sampling}
\bibfield{author}{\bibinfo{person}{J. Chen}, \bibinfo{person}{V. Nair},
  \bibinfo{person}{R. Krishna}, {and} \bibinfo{person}{T. Menzies}.}
  \bibinfo{year}{2016}\natexlab{}.
\newblock \showarticletitle{Is Sampling better than Evolution for Search-based
  Software Engineering?}
\newblock \bibinfo{journal}{{\em arXiv\/}} (\bibinfo{year}{2016}).
\newblock


\bibitem[\protect\citeauthoryear{Efron and Tibshirani}{Efron and
  Tibshirani}{1993}]%
        {efron93}
\bibfield{author}{\bibinfo{person}{B. Efron} {and} \bibinfo{person}{R.~J.
  Tibshirani}.} \bibinfo{year}{1993}\natexlab{}.
\newblock \bibinfo{booktitle}{{\em {An Introduction to the Bootstrap}}}.
\newblock \bibinfo{publisher}{CRC}.
\newblock


\bibitem[\protect\citeauthoryear{Foss, Stensrud, Kitchenham, and Myrtveit}{Foss
  et~al\mbox{.}}{2003}]%
        {foss2003simulation}
\bibfield{author}{\bibinfo{person}{T. Foss}, \bibinfo{person}{E. Stensrud},
  \bibinfo{person}{B. Kitchenham}, {and} \bibinfo{person}{I. Myrtveit}.}
  \bibinfo{year}{2003}\natexlab{}.
\newblock \showarticletitle{A Simulation Study of the Model Evaluation
  Criterion MMRE}.
\newblock \bibinfo{journal}{{\em IEEE Transactions on Software Engineering
  (TSE)\/}}  \bibinfo{volume}{29} (\bibinfo{year}{2003}),
  \bibinfo{pages}{985--995}.
\newblock


\bibitem[\protect\citeauthoryear{Ghotra, McIntosh, and Hassan}{Ghotra
  et~al\mbox{.}}{2015}]%
        {7194626}
\bibfield{author}{\bibinfo{person}{B. Ghotra}, \bibinfo{person}{S. McIntosh},
  {and} \bibinfo{person}{A.~E. Hassan}.} \bibinfo{year}{2015}\natexlab{}.
\newblock \showarticletitle{Revisiting the Impact of Classification Techniques
  on the Performance of Defect Prediction Models}. In \bibinfo{booktitle}{{\em
  Proc. of International Conference on Software Engineering (ICSE)}}. IEEE,
  \bibinfo{pages}{789--800}.
\newblock


\bibitem[\protect\citeauthoryear{Guo, Czarnecki, Apel, Siegmund, and
  Wasowski}{Guo et~al\mbox{.}}{2013}]%
        {guo2013variability}
\bibfield{author}{\bibinfo{person}{J. Guo}, \bibinfo{person}{K. Czarnecki},
  \bibinfo{person}{S. Apel}, \bibinfo{person}{N. Siegmund}, {and}
  \bibinfo{person}{A. Wasowski}.} \bibinfo{year}{2013}\natexlab{}.
\newblock \showarticletitle{Variability-Aware Performance Prediction: A
  Statistical Learning Approach}. In \bibinfo{booktitle}{{\em Proc. of
  International Conference on Automated Software Engineering (ASE)}}. IEEE,
  \bibinfo{pages}{301--311}.
\newblock


\bibitem[\protect\citeauthoryear{Henard, Papadakis, Harman, and Traon}{Henard
  et~al\mbox{.}}{2015}]%
        {henard2015combining}
\bibfield{author}{\bibinfo{person}{C. Henard}, \bibinfo{person}{M. Papadakis},
  \bibinfo{person}{M. Harman}, {and} \bibinfo{person}{Y.~Le Traon}.}
  \bibinfo{year}{2015}\natexlab{}.
\newblock \showarticletitle{Combining Multi-Objective Search and Constraint
  Solving for Configuring Large Software Product Lines}. In
  \bibinfo{booktitle}{{\em Proc. of International Conference on Software
  Engineering (ICSE)}}. IEEE, \bibinfo{pages}{517--528}.
\newblock


\bibitem[\protect\citeauthoryear{Jamshidi and Casale}{Jamshidi and
  Casale}{2016}]%
        {jamshidi2016uncertainty}
\bibfield{author}{\bibinfo{person}{P. Jamshidi} {and} \bibinfo{person}{G.
  Casale}.} \bibinfo{year}{2016}\natexlab{}.
\newblock \showarticletitle{An Uncertainty-Aware Approach to Optimal
  Configuration of Stream Processing Systems}. In \bibinfo{booktitle}{{\em
  Proc. of International Symposium on Modeling, Analysis and Simulation of
  Computer and Telecommunication Systems (MASCOTS)}}. IEEE,
  \bibinfo{pages}{39--48}.
\newblock


\bibitem[\protect\citeauthoryear{Kloke and McKean}{Kloke and McKean}{2012}]%
        {kloke2012rfit}
\bibfield{author}{\bibinfo{person}{J.D. Kloke} {and} \bibinfo{person}{J.~W.
  McKean}.} \bibinfo{year}{2012}\natexlab{}.
\newblock \showarticletitle{Rfit: Rank-based Estimation for Linear Models}.
\newblock \bibinfo{journal}{{\em The R Journal\/}}  \bibinfo{volume}{4}
  (\bibinfo{year}{2012}), \bibinfo{pages}{57--64}.
\newblock


\bibitem[\protect\citeauthoryear{Krall, Menzies, and Davies}{Krall
  et~al\mbox{.}}{2015}]%
        {krall2015gale}
\bibfield{author}{\bibinfo{person}{J. Krall}, \bibinfo{person}{T. Menzies},
  {and} \bibinfo{person}{M. Davies}.} \bibinfo{year}{2015}\natexlab{}.
\newblock \showarticletitle{GALE: Geometric Active Learning for Search-Based
  Software Engineering}.
\newblock \bibinfo{journal}{{\em IEEE Transactions on Software Engineering
  (TSE)\/}}  \bibinfo{volume}{41} (\bibinfo{year}{2015}),
  \bibinfo{pages}{1001--1018}.
\newblock


\bibitem[\protect\citeauthoryear{Lim, Jin, Ong, and Sendhoff}{Lim
  et~al\mbox{.}}{2010}]%
        {lim2010generalizing}
\bibfield{author}{\bibinfo{person}{D. Lim}, \bibinfo{person}{Y. Jin},
  \bibinfo{person}{Y.~S. Ong}, {and} \bibinfo{person}{B. Sendhoff}.}
  \bibinfo{year}{2010}\natexlab{}.
\newblock \showarticletitle{Generalizing Surrogate-Assisted Evolutionary
  Computation}.
\newblock \bibinfo{journal}{{\em IEEE Transactions on Evolutionary
  Computation\/}}  \bibinfo{volume}{14} (\bibinfo{year}{2010}),
  \bibinfo{pages}{329--355}.
\newblock


\bibitem[\protect\citeauthoryear{Menzies, Butcher, Cok, Marcus, Layman, Shull,
  Turhan, and Zimmermann}{Menzies et~al\mbox{.}}{2013}]%
        {me12d}
\bibfield{author}{\bibinfo{person}{T. Menzies}, \bibinfo{person}{A. Butcher},
  \bibinfo{person}{D. Cok}, \bibinfo{person}{A. Marcus}, \bibinfo{person}{L.
  Layman}, \bibinfo{person}{F. Shull}, \bibinfo{person}{B. Turhan}, {and}
  \bibinfo{person}{T. Zimmermann}.} \bibinfo{year}{2013}\natexlab{}.
\newblock \showarticletitle{Local versus Global Lessons for Defect Prediction
  and Effort Estimation}.
\newblock \bibinfo{journal}{{\em IEEE Transactions on Software Engineering
  (TSE)\/}}  \bibinfo{volume}{39} (\bibinfo{year}{2013}),
  \bibinfo{pages}{822--834}.
\newblock


\bibitem[\protect\citeauthoryear{Mittas and Angelis}{Mittas and
  Angelis}{2013}]%
        {mittas13}
\bibfield{author}{\bibinfo{person}{N. Mittas} {and} \bibinfo{person}{L.
  Angelis}.} \bibinfo{year}{2013}\natexlab{}.
\newblock \showarticletitle{Ranking and Clustering Software Cost Estimation
  Models through a Multiple Comparisons Algorithm}. \bibinfo{journal}{{\em IEEE
  Transactions on Software Engineering (TSE)\/}}  \bibinfo{volume}{39}
  (\bibinfo{year}{2013}), \bibinfo{pages}{537--551}.
\newblock


\bibitem[\protect\citeauthoryear{Myrtveit and Stensrud}{Myrtveit and
  Stensrud}{2012}]%
        {myrtveit2012validity}
\bibfield{author}{\bibinfo{person}{I. Myrtveit} {and} \bibinfo{person}{E.
  Stensrud}.} \bibinfo{year}{2012}\natexlab{}.
\newblock \showarticletitle{Validity and Reliability of Evaluation Procedures
  in Comparative Studies of Effort Prediction Models}.
\newblock \bibinfo{journal}{{\em Empirical Software Engineering (ESE)\/}}
  \bibinfo{volume}{17} (\bibinfo{year}{2012}), \bibinfo{pages}{23--33}.
\newblock


\bibitem[\protect\citeauthoryear{Myrtveit, Stensrud, and Shepperd}{Myrtveit
  et~al\mbox{.}}{2005}]%
        {myrtveit2005reliability}
\bibfield{author}{\bibinfo{person}{I. Myrtveit}, \bibinfo{person}{E. Stensrud},
  {and} \bibinfo{person}{M. Shepperd}.} \bibinfo{year}{2005}\natexlab{}.
\newblock \showarticletitle{Reliability and Validity in Comparative Studies of
  Software Prediction Models}.
\newblock \bibinfo{journal}{{\em IEEE Transactions on Software Engineering
  (TSE)\/}}  \bibinfo{volume}{31} (\bibinfo{year}{2005}),
  \bibinfo{pages}{380--391}.
\newblock


\bibitem[\protect\citeauthoryear{Nair, Menzies, Siegmund, and Apel}{Nair
  et~al\mbox{.}}{2017}]%
        {nair2017faster}
\bibfield{author}{\bibinfo{person}{V. Nair}, \bibinfo{person}{T. Menzies},
  \bibinfo{person}{N. Siegmund}, {and} \bibinfo{person}{S. Apel}.}
  \bibinfo{year}{2017}\natexlab{}.
\newblock \showarticletitle{Faster Discovery of Faster System Configurations
  with Spectral Learning}.
\newblock \bibinfo{journal}{{\em arXiv\/}} (\bibinfo{year}{2017}).
\newblock


\bibitem[\protect\citeauthoryear{Pedregosa, Varoquaux, Gramfort, Michel,
  Thirion, Grisel, Blondel, Prettenhofer, Weiss, Dubourg,
  et~al\mbox{.}}{Pedregosa et~al\mbox{.}}{2011}]%
        {scikit-learn}
\bibfield{author}{\bibinfo{person}{F. Pedregosa}, \bibinfo{person}{G.
  Varoquaux}, \bibinfo{person}{A. Gramfort}, \bibinfo{person}{V. Michel},
  \bibinfo{person}{B. Thirion}, \bibinfo{person}{O. Grisel},
  \bibinfo{person}{M. Blondel}, \bibinfo{person}{P. Prettenhofer},
  \bibinfo{person}{R. Weiss}, \bibinfo{person}{V. Dubourg}, {and}
  \bibinfo{person}{others}.} \bibinfo{year}{2011}\natexlab{}.
\newblock \showarticletitle{Scikit-learn: Machine Learning in Python}.
\newblock \bibinfo{journal}{{\em Journal of Machine Learning Research\/}}
  \bibinfo{volume}{12} (\bibinfo{year}{2011}), \bibinfo{pages}{2825--2830}.
\newblock


\bibitem[\protect\citeauthoryear{Rosset, Perlich, and Zadrozny}{Rosset
  et~al\mbox{.}}{2005}]%
        {rosset2005ranking}
\bibfield{author}{\bibinfo{person}{S. Rosset}, \bibinfo{person}{C. Perlich},
  {and} \bibinfo{person}{B. Zadrozny}.} \bibinfo{year}{2005}\natexlab{}.
\newblock \showarticletitle{Ranking-based Evaluation of Regression Models}. In
  \bibinfo{booktitle}{{\em Proc. of International Conference on Data Mining
  (ICDM)}}. IEEE.
\newblock


\bibitem[\protect\citeauthoryear{Sarkar, Guo, Siegmund, Apel, and
  Czarnecki}{Sarkar et~al\mbox{.}}{2015}]%
        {sarkar2015cost}
\bibfield{author}{\bibinfo{person}{A. Sarkar}, \bibinfo{person}{J. Guo},
  \bibinfo{person}{N. Siegmund}, \bibinfo{person}{S. Apel}, {and}
  \bibinfo{person}{K. Czarnecki}.} \bibinfo{year}{2015}\natexlab{}.
\newblock \showarticletitle{Cost-Efficient Sampling for Performance Prediction
  of Configurable Systems (T)}. In \bibinfo{booktitle}{{\em Proc. of
  International Conference on Automated Software Engineering (ASE)}}. IEEE,
  \bibinfo{pages}{342--352}.
\newblock


\bibitem[\protect\citeauthoryear{Sayyad, Ingram, Menzies, and Ammar}{Sayyad
  et~al\mbox{.}}{2013}]%
        {sayyad2013scalable}
\bibfield{author}{\bibinfo{person}{A.~S. Sayyad}, \bibinfo{person}{J. Ingram},
  \bibinfo{person}{T. Menzies}, {and} \bibinfo{person}{H. Ammar}.}
  \bibinfo{year}{2013}\natexlab{}.
\newblock \showarticletitle{Scalable Product Line Configuration: A Straw to
  Break the Camel's Back}. In \bibinfo{booktitle}{{\em Proc. of International
  Conference on Automated Software Engineering (ASE)}}. IEEE,
  \bibinfo{pages}{465--474}.
\newblock


\bibitem[\protect\citeauthoryear{Siegmund, Siegmund, and Apel}{Siegmund
  et~al\mbox{.}}{2015}]%
        {SSA15}
\bibfield{author}{\bibinfo{person}{J. Siegmund}, \bibinfo{person}{N. Siegmund},
  {and} \bibinfo{person}{S. Apel}.} \bibinfo{year}{2015}\natexlab{}.
\newblock \showarticletitle{Views on Internal and External Validity in
  Empirical Software Engineering}. In \bibinfo{booktitle}{{\em Proc. of
  International Conference on Software Engineering (ICSE)}}. IEEE,
  \bibinfo{pages}{9--19}.
\newblock


\bibitem[\protect\citeauthoryear{Siegmund, Kolesnikov, K\"astner, Apel, Batory,
  Rosenm{\"u}ller, and Saake}{Siegmund et~al\mbox{.}}{2012}]%
        {siegmund2012predicting}
\bibfield{author}{\bibinfo{person}{N. Siegmund}, \bibinfo{person}{S.~S.
  Kolesnikov}, \bibinfo{person}{C. K\"astner}, \bibinfo{person}{S. Apel},
  \bibinfo{person}{D. Batory}, \bibinfo{person}{M. Rosenm{\"u}ller}, {and}
  \bibinfo{person}{G. Saake}.} \bibinfo{year}{2012}\natexlab{}.
\newblock \showarticletitle{Predicting Performance via Automated
  Feature-Interaction Detection}. In \bibinfo{booktitle}{{\em Proc. of
  International Conference on Software Engineering (ICSE)}}. IEEE,
  \bibinfo{pages}{167--177}.
\newblock


\bibitem[\protect\citeauthoryear{Vargha and Delaney}{Vargha and
  Delaney}{2000}]%
        {Vargha00}
\bibfield{author}{\bibinfo{person}{A. Vargha} {and} \bibinfo{person}{H.~D.
  Delaney}.} \bibinfo{year}{2000}\natexlab{}.
\newblock \showarticletitle{A Critique and Improvement of the CL Common
  Language Effect Size Statistics of McGraw and Wong}. \bibinfo{journal}{{\em
  Journal of Educational and Behavioral Statistics\/}}  \bibinfo{volume}{25}
  (\bibinfo{year}{2000}), \bibinfo{pages}{101--132}.
\newblock


\bibitem[\protect\citeauthoryear{Weiss and Tian}{Weiss and Tian}{2008}]%
        {weiss2008maximizing}
\bibfield{author}{\bibinfo{person}{G.~M. Weiss} {and} \bibinfo{person}{Y.
  Tian}.} \bibinfo{year}{2008}\natexlab{}.
\newblock \showarticletitle{Maximizing Classifier Utility when there are Data
  Acquisition and Modeling Costs}.
\newblock \bibinfo{journal}{{\em Journal of Data Mining and Knowledge
  Discovery\/}} (\bibinfo{year}{2008}), \bibinfo{pages}{253--282}.
\newblock


\bibitem[\protect\citeauthoryear{Xu, Jin, Fan, Zhou, Pasupathy, and
  Talwadker}{Xu et~al\mbox{.}}{2015}]%
        {xu2015hey}
\bibfield{author}{\bibinfo{person}{T. Xu}, \bibinfo{person}{L. Jin},
  \bibinfo{person}{X. Fan}, \bibinfo{person}{Y. Zhou}, \bibinfo{person}{S.
  Pasupathy}, {and} \bibinfo{person}{R. Talwadker}.}
  \bibinfo{year}{2015}\natexlab{}.
\newblock \showarticletitle{Hey, You Have Given Me Too Many Knobs!:
  Understanding and Dealing with Over-designed Configuration in System
  Software}. In \bibinfo{booktitle}{{\em Proc. of Joint Meeting on Foundations
  of Software Engineering (ESEC/FSE)}}. ACM, \bibinfo{publisher}{ACM},
  \bibinfo{pages}{307--319}.
\newblock


\end{thebibliography}
